\documentclass[preprint]{aastex}
\usepackage{emulateapj5}
\def\stacksymbols #1#2#3#4{\def\theguybelow{#2}
	\def\verticalposition{\lower#3pt}
	\def\spacingwithinsymbol{\baselineskip0pt\lineskip#4pt}
	\mathrel{\mathpalette\intermediary#1}}
\def\intermediary #1#2{\verticalposition\vbox{\spacingwithinsymbol
	\everycr={}\tabskip0pt
	\halign{$\mathsurround0pt#1\hfil##\hfil$\crcr#2\crcr
		\theguybelow\crcr}}}
\def\lta{\stacksymbols{<}{\sim}{2.5}{.2}}
\def\gta{\stacksymbols{>}{\sim}{3}{.5}}

\begin{document}

\title{FEEDBACK HEATING IN CLUSTER AND GALACTIC COOLING FLOWS}

\author{Fabrizio Brighenti$^{1,2}$ \& William G. Mathews$^1$}

\affil{$^1$University of California Observatories/Lick Observatory,
Board of Studies in Astronomy and Astrophysics,
University of California, Santa Cruz, CA 95064\\
mathews@lick.ucsc.edu}

\affil{$^2$Dipartimento di Astronomia,
Universit\`a di Bologna,
via Ranzani 1,
Bologna 40127, Italy\\
brighenti@bo.astro.it}

%\vskip 2.in
%\noindent
%Received:

%\noindent
%PROOFS TO BE SENT TO:

%\noindent
%Lick Observatory

%\noindent
%Santa Cruz, CA 95064

%\noindent
%$^1$UCO/Lick Observatory Bulletin No.

%\vskip3.in
%\noindent
%Short Title: X-ray Holes in Cooling Flows 
%\clearpage
\vskip .2in

\begin{abstract}
Cluster cooling flow models that include both AGN heating 
and thermal conduction have lower overall mass cooling 
rates and simultaneously sustain density and temperature 
profiles similar to those observed. 
These computed flows have no {\it ad hoc} mass dropout.
To achieve this agreement, the thermal conductivity must 
be about $0.35 \pm 0.10$ of the Spitzer value, similar to 
that advocated by Narayan \& Medvedev. 
However, when applied to galaxy/group scales 
the synergistic 
combination of AGN heating and conduction is less satisfactory. 
When the computed density profile 
and the global cooling rate are lowered 
by AGN heating to match observations of these smaller scale 
flows, 
the gas temperatures within $\sim 10$ kpc are too large. 
In addition, best-fitting 
flows in galaxy/groups with AGN heating and thermal conduction 
require conductivities much closer 
to the Spitzer value $\sim 0.5 - 1$. 
Another difficulty with galaxy/group flows that combine AGN 
heating and conduction is that the iron enrichment 
by Type Ia supernovae is more 
effective when the gas density is lowered by heating to 
match the observations. 
The hot gas iron abundance in galactic flows with heating 
and conduction greatly exceeds 
observed values throughout most of the galaxy.
Galactic/group flows with central 
heating and conduction therefore require 
an additional process that removes the iron: 
failure of Type Ia ejecta to go into the hot phase, 
selective cooling, etc.
\end{abstract}


%%%%%%%%\clearpage

\section{Introduction}

Recent Chandra and XMM observations of hot gas in clusters and 
galaxies have radically altered our previous models for cooling 
flows and presented new possibilities for understanding them 
in a different way.
X-ray spectra taken with the XMM RGS (Reflection Grating Spectrometer) 
fail to show line emission from ions having intermediate 
or low temperatures, 
implying that the cooling gas is somehow hidden from view or that 
the cooling rate is at least 5 or 10 times less than 
previously assumed (e.g. Peterson et al. 2001; Tamura et al. 2001;
Kaastra et al. 2001; Xu et al. 2001; 
Sakelliou, et al. 2002). XMM-EPIC observations also fail to detect
cooling gas (Molendi \& Pizzolato 2001; B\"ohringer et al. 2002).
However, X-ray images reveal that the 
hot gas inside the E or cD galaxies located at 
the centers of cooling flows is often highly disturbed 
(e.g. B\"ohringer et al. 1993; Fabian et al 2000; 
McNamara et al. 2000; Loewenstein et al. 2000;
Blanton et al 2001;
Jones et al. 2002; 
Forman et al. 2001; Buote et al. 2002; 
Trinchieri \& Goudfrooij 2002). 
Evidently, massive black holes, thought to 
lie at the centers of all stellar bulges, become intermittently 
active possibly stimulated by the inflow of hot gas.
Of particular interest are the cavities in the hot gas 
that are often (but not always) associated with radio lobes
as in Perseus/NGC 1275 (Fabian et al 2000) 
and HydraA/3C295 (McNamara et al. 2000; David et al. 2001;
Allen et al 2001).
Cavities occur when ultrahot or relativistic gas displaces 
the hot thermal gas.
Since the rims surrounding the X-ray cavities are 
typically no hotter than other ambient gas
(Fabian 2001; McNamara 2001), 
strong shocks do not seem to be involved in producing the holes 
so the innerhole gas is 
in approximate pressure equilibrium with gas in the rims 
(Nulsen et al 2002; Soker, Blanton \& Sarazin 2002; 
Brighenti \& Mathews 2002c).
Obviously the holes must be buoyant (Churazov et al. 2001) 
and their formation and subsequent evolution must feed energy into 
the cooling flow gas. 

Can the heating visible in Chandra images explain the absence of 
cooling in XMM spectra? 
This question can be answered only from studies of the global 
effect of many heating episodes over many Gyrs. 
Recently we presented a number of evolutionary models of 
so-called cooling flows that were heated by a variety of scenarios 
initiated by gas flowing into the central black hole 
(Brighenti \& Mathews 2002b).
None of the flow models we considered were satisfactory; 
whenever the heating was sufficient to reduce the rate that gas 
cools near the center, the flow was highly disrupted and 
spatially distributed cooling occurred at larger radii at essentially 
the same rate as in the unheated flows.

In the light of several recent developments -- including the
significant continued interest in heated flows -- 
we are encouraged to reconsider such models again here.
Our first motivation is 
the argument by Narayan \& Medvedev (2001) 
that thermal conductivity 
is reduced only by $\sim 0.2$ in the presence of tangled 
magnetic fields which are expected in cooling flows. 
Secondly, in a recent paper Ruszkowski \& Begelman (2002) 
demonstrated that heating and thermal conduction 
can conspire to produce successful cluster 
cooling flows that have the 
characteristic positive temperature gradient at small radii, 
a common observed feature in all these flows.
Finally, in our models we follow the prescription of 
Ruszkowski \& Begelman and other authors
(e.g. Binney \& Tabor 1995; Tucker \& David 1997; 
Ciotti \& Ostriker 2001)
by considering cluster and galactic 
flows in which the mechanical energy released by the 
central AGN is exactly proportional to the 
mass of gas that flows into the center of the flow, 
optimizing the effect of the energy feedback.

In the following models we consider the long term evolution of 
hot gas on both cluster and galactic scales but with different 
approaches. 
Our models are intended to be generic, but as a guide to their 
success we compare them with recent observations of 
Abell 1795 and NGC 4472 for cluster and galactic flows respectively.
Since cluster flows are occasionally  disturbed by merging events, 
we begin with a quiescent post-merger 
flow initially in hydrostatic equilibrium 
and follow its evolution for several Gyrs under the influence 
of radiative cooling, AGN feedback 
heating in the core, and thermal conduction.
For galactic flows, we begin with a model elliptical having an 
additional complement of circumgalactic hot gas and follow its  
evolution for 12 Gyrs under the influence of radiative cooling, 
AGN heating and conduction as well as the additional effects
of Type Ia supernovae and stellar mass loss appropriate to 
galactic flows.
We find that it is possible to get marginally 
satisfactory solutions 
for cluster flows provided the parameters are chosen within 
rather narrow limits.
However, our models of galactic cooling flows with the same 
choice of parameters are inadequate or require additional 
possibly unsatisfactory assumptions.

\section{Computational Methods}

The calculations described below are based on the same flow equations 
described in detail in our previous work on heated cooling flows 
(Brighenti \& Mathews 2002b) and the reader is referred to that paper 
for details.
We assume no distributed radiative cooling, but this may occur 
naturally in convective regions when mass elements collide and 
compress. 
In such instances we remove the gas from the flow 
when the temperature has dropped to $\sim T_{cool} = 5 \times 10^5$ K 
as described in Brighenti \& Mathews (2002b).
Results of  
1D and 2D are in essential 
agreement when applied to the same flow 
(see Brighenti \& Mathews 2002b) although 2D flows 
are essential to describe flows with off-center heating.

The grid spacing we use varies with the overall scale of the 
flow and the number of dimensions considered.
{\it 2D Cluster}: For these flows we use 
$250 \times 250$ cylindrical zones. 
The grid is uniform at 2.5 kpc spacing 
for the inner $150 \times 150$ zones 
extending to $375 \times 375$ kpc. 
Beyond this region the zone size increases geometrically 
to 2 Mpc. 
We find very similar results with a $490 \times 490$ grid 
where the inner $300 \times 300$ zones are 0.4 kpc wide.
{\it 1D Cluster}: These calculations are done with 
360 zones increasing in size from 
0.5 kpc at the center to an outer boundary at 3 Mpc.
{\it 2D Galaxy/Group}: These flows are 
calculated with $280 \times 280$ cylindrical zones. 
The grid is uniform at 0.25 kpc spacing
for the inner $200 \times 200$ zones 
that extend to $50 \times 50$ kpc. 
Beyond this region the zone size increases geometrically 
to 1.7 Mpc. 
{\it 1D Galaxy/Group}: Here we use 
350 zones increasing in size from 
0.05 kpc at the center to an outer boundary at 1.6 Mpc.

\section{Cluster Flows with AGN Heating, Cooling and Conduction}

Our cluster flow calculations begin in hydrostatic 
equilibrium with temperature and density 
profiles based on the well-observed cluster Abell 1795 
assumed to be at a distance of 243 Mpc.
Abell 1795 is a typical rich cluster 
with a central cD galaxy and reasonably 
relaxed overall structure (Boute \& Tsai 1996). 
Abell 1795 
has the usual attributes of normal cooling flows:
strong central peak in X-ray surface brightness
(e.g. Tamura et al. 2001),
a central radiative cooling time 
$\sim 3 \times 10^8$ yrs that is much less than 
the cluster age (e.g. Edge et al. 1992; Fabian et al. 2001), 
optical line emission near the central cD 
(Cowie et al. 1983),
an excess of blue and ultraviolet light from 
massive young stars (Johnstone, Fabian \& Nulsen 1987;
Cardiel, Gorgas \& Aragon-Salamanca 1998; Mittaz et a. 2001)
and a central radio source 4C 26.42 (McNamara et al. 1996a,b).
Near the center of Abell 1795 Chandra images reveal   
an X-ray emission feature aligned with an optical 
filament (Fabian et al. 2001). 
This and the central total mass profile,
$M \propto r^{0.6}$ inside 40 kpc, which is somewhat flatter 
than NFW, suggest a possible deviation from hydrostatic 
equilibrium.
Deprojections of ROSAT images of Abell 1795 led 
Allen et al. (2000) to conclude that the total cooling 
rate is ${\dot M} \sim 500$ $M_{\odot}$ yr$^{-1}$. 
However, the XMM RGS spectrum showed no evidence of gas 
with temperatures less than $\sim 2$ keV, 
corresponding to a much smaller upper limit to the cooling 
rate, ${\dot M} < 150$ $M_{\odot}$ yr$^{-1}$.
This is consistent with a rate 
${\dot M} \approx 100$ $M_{\odot}$ yr$^{-1}$ estimated from 
Chandra observations (Ettori et al. 2002).

In the upper two rows of Figure 1 we show 
electron density and temperature 
profiles of Abell 1795 determined 
from observations with 
XMM (filled triangles; Tamura et al. 2001) 
and Chandra (open triangles; Ettori et al. 2002).
The Chandra temperatures have been spatially deprojected and 
are therefore slightly lower.
We inserted the observed density and temperature profiles 
into the equation of hydrostatic equilibrium to determine the 
mass distribution. 
In this process we ignore the possible deviations from 
equilibrium suggested by the Chandra observations. 
Nevertheless, our results described below 
are independent of the particular mass 
profile assumed which could have been strictly NFW for example.
Our adopted fits $n(r)$ and $T(r)$ to the observed profiles 
are shown as dot-dashed lines in the first column of Figure 1. 
The central density and temperature in these fits are 
$n(0) = 0.1$ cm$^{-3}$ and $T(0) = 2.5 \times 10^7$ K. 

\subsection{Unheated Cluster Flow}

In the first column of Figure 1 we illustrate the evolution 
of Abell 1795 as determined by radiative losses only  
with no AGN or conductive heating. 
Our calculations are done in 2D and the results in Figure 1 are 
azimuthally averaged. 
For standard cosmological parameters 
($\Omega = 0.3$; $\Lambda = 0.7$; $H = 70$ km s$^{-1}$ Mpc$^{-1}$) 
large clusters like Abell 1795 were formed only recently,
so we consider their evolution for only 7 Gyrs.
As gas within $\sim 30$ kpc cools, 
a subsonic inflow develops to reestablish approximate equilibrium, 
forcing the central density to rise above the observations.
In the bottom row of Figure 1 we plot the evolution 
of the cooling rate ${\dot M}(t)$  
which increases from 
${\dot M} \approx 100$ $M_{\odot}$ yr$^{-1}$ at the 
beginning of the calculation toward an asymptotic 
value of $\sim 300$ $M_{\odot}$ yr$^{-1}$
after $\sim 4$ Gyrs 
when a quasi-steady inflow is established.
Not only does this ${\dot M}(t)$ exceed the cooling rate 
allowed by XMM spectra, 
all the cooling occurs at the very center where 
$\sim 3 \times 10^{11}$ $M_{\odot}$ accumulates every Gyr. 
Such an unacceptably 
huge mass would double the mass of the central cD in a few Gyrs.
The results of each evolutionary calculation are 
summarized in Table 1.

\subsection{Cluster Flows with Heating}

In the next series of models we consider the effect of 
central and off-central AGN heating. 
The heating power $L_h = \varepsilon {\dot M}(0) c^2$ erg s$^{-1}$
is proportional to the rate ${\dot M}(0)$ that gas flows into 
the central grid zone.
The efficiency $\varepsilon$ is a combined 
measure of both the AGN power generated and the fraction that 
is delivered to the hot cluster gas.
As in our previous paper (Brighenti \& Mathews 2002b) we assume 
that the heating is instantaneously transferred to the hot 
gas near the flow center and that the degree of heating has 
a Gaussian profile $\propto \exp[-(r/r_h)^2]$. 
The spatial 
scale of the heated region $r_h$ has only a secondary influence 
on our results; we restrict our discussion here to $r_h = 25$ kpc. 
Our two dimensional (2D) calculations of heated flows 
create buoyant bubbles that transport energy to larger radii 
where they eventually dissolve into the background.
The merging of the hot bubbles with the background flow 
is probably an artifact of numerical diffusion in our code,
but a similar dissipation may be naturally accomplished 
by thermal conduction at the bubble surfaces, 
where the surface area is greatly 
enlarged due to Rayleigh-Taylor instabilities.
We do not attempt to follow the detailed 
bubble physics here, but only ensure that the 
injected energy is globally conserved.

The second and third columns in Figure 1 shows the consequences 
of Gaussian feedback 
heating on the 2D flow in the first column using 
$\varepsilon = 10^{-4}$ and $10^{-3}$ respectively.
When $\varepsilon < 10^{-4}$, the flow evolves in a manner 
similar to the unheated flow in the first column.
Irregularities in the density and temperature profiles are 
introduced by the heating intermittancy. 
With $\varepsilon = 10^{-4}$ the density and temperature 
profiles after 4 and 7 Gyrs are almost acceptable, but 
${\dot M}$ (third row of Fig. 1) is too large, 
$\gta 150$ $M_{\odot}$ yr$^{-1}$ beyond 4.5 Gyrs. 
When the heating efficiency is increased to 
$\varepsilon = 10^{-3}$ (third column of Fig. 1), 
the central cooling rate, 
${\dot M} < 20$ $M_{\odot}$ yr$^{-1}$, is acceptable.
However, for this higher efficiency 
the gas is much too hot within the heated region 
and resembles no known cluster. 

The results of similar evolutionary calculations 
with off-center heating (similar to a one-sided jet) are shown 
in the final two columns of Figure 1. 
The Gaussian heating is displaced 50 kpc 
from the flow center along the $z$-axis, 
$(z,R)_h = (50~{\rm kpc},0~{\rm kpc})$, 
but all other parameters are 
identical to the centrally heated flows.
Figure 1 shows azimuthally averaged profiles 
of 2D calculations. 
The $\varepsilon = 10^{-4}$ case has a marginally acceptable 
central cooling rate, 
${\dot M} \lta 30$ $M_{\odot}$ yr$^{-1}$, but the 
gas temperature profile within $\sim 100$ kpc is flatter 
than normal. 
As before $\varepsilon = 10^{-3}$ lowers ${\dot M}$ 
to acceptable values but clearly overheats the gas.

In summary, both heated and unheated flows deviate from 
typical cluster profiles after a few Gyrs. 
When the heating is large enough to satisfy the upper limits
on ${\dot M}$ imposed by XMM spectra, 
the flow profiles are too distorted to be acceptable.  
There is only a very small range of heating efficiencies 
for which these problems are minimized (but not removed). 
These general conclusions also apply to similar calculations 
with different heating scales $r_h$.
Finally,
all the gas cools at or near the origin in these heated flows 
(Table 1).

\subsection{Cluster Flows with Conduction and Heating}

Since the temperature gradient is negative throughout 
the observed region in Abell 1795,  
the central parts of the flow are 
conductively heated from the outside.
However, just after the center of the flow is heated with 
AGN energy, thermal conduction rapidly transfers this 
energy outward toward larger radii.
We first discuss the influence of conduction alone and 
then the combined effect of conduction 
plus heating on the evolution of Abell 1795. 
Because of the shorter computational time steps when conduction 
is included, we describe the evolution of these flows in 1D. 
Based on examples discussed in our previous paper 
(Brighenti \& Mathews 2002b),
we are confident that the 1D calculations are trustworthy 
since 1D and 2D models are in essential agreement in every 
case that we have tested. 

Thermal conductivity in a hot plasma,
$\kappa = 1.84 \times 10^{-5} (\ln \Lambda)^{-1} T^{5/2}$
erg/sec cm K  can be important at
high temperatures, but is reduced 
by magnetic fields.
The effect of magnetic suppression on the thermal flux 
$F_{cond} = f\kappa (dT/dr)$ is represented with a coefficient 
$f \le 1$ which is assumed to be uniform in the flow. 
It is well known that thermal conduction perpendicular to 
a uniform magnetic field is lowered by many orders of 
magnitude, 
so it has generally been assumed that $f \ll 1$ applied 
to any plasma with a reasonably tangled magnetic field. 
However, Narayan \& Medvedev (2001) have recently shown that 
$f \sim 0.2$ is appropriate for 
thermal conduction in a hot plasma with
chaotic magnetic field fluctuations and this opens up 
new possibilities for cooling flows.

The evolution of Abell 1795 with thermal conduction but 
no AGN heating is illustrated in the first column of Figure 2. 
Each curve in Figure 2 is labeled with three parameters, 
($\varepsilon$,$f$,$t$ in Gyrs).
The influence of conduction on unheated cluster flows 
is quite sensitive to $f$. 
When $f = 0.5$ the conductive flux is so high that the 
solution begins to deviate from the observations after 
only 1 Gyr and by 4 Gyrs the  
temperature profile is almost isothermal. 
This $T(r)$ is clearly discordant 
with observations 
although ${\dot M}$ is nicely suppressed (Table 1). 
The flow with $f = 0.25$ is similar to that of the unheated 
flow in Figure 1.
The value of $f$ required to maintain the observed profiles 
in Abell 1795 is $0.35 \pm 0.10$
(see Table 1), but this optimal value probably 
varies from cluster to cluster.
The fine-tuning of the $f$ parameter 
as well as the difficulty in balancing radiative losses 
near the centers of cooling flows with the inward conduction 
of heat from larger radii are well known from previous studies 
(Stewart, Canizares, Fabian \& Nulsen 1984;
Bertschinger \& Meiksin 1986; Meiksin 1988;
Bregman \& David 1988;
Loewenstein, Zweibel \& Begelman 1991;
Brighenti \& Mathews 2002b).

However, the effect of thermal conduction in heated flows helps to
spread the AGN heating into the flow, preserving smooth temperature
profiles resembling the observations.  Flows with AGN heating and
conduction are shown in the second and third columns of Figure 2.  For
$f = 0.25$ and $10^{-4} \lta \varepsilon \lta 10^{-3}$ (second column
of Fig. 2) both the density and temperature profiles are maintained
for up to 4 Gyrs.  These results are similar to the model described by
Ruszkowski \& Begelman (2002) in which a cluster flow is heated by
conduction from the outside and the entire flow is instantaneously
heated from the inside when gas flows into the central AGN.  We have
duplicated the Ruszkowski-Begelman cluster flow with our own program
using their type of AGN heating.  In general a combination of central
AGN heating and thermal conduction can produce cluster flows with many
desirable features.  While the details of the flows depend on the
spatial scale $r_h$ of the centrally heated regions, satisfactory
gasdynamic flows in general agreement with cluster observations can be
found for a wide range of feedback heating scales $r_h$.  
However,
unlike Ruszkowski \& Begelman, we find that the results are sensitive
to the value of the magnetic suppression coefficient $f$, as
illustrated in the third column of Figure 2.  
For a given heating
efficiency $\varepsilon$ the flow evolves toward isothermality if $f =
0.5$ and cools similar to the unheated flow (column 1, Fig. 1) if $f =
0.1$.  
The ${\dot M}$ for the $f = 0.1$ solution also exceeds the observed 
limits for $t > 4$ Gyrs.
Happily, these bad outcomes bracket the value $f = 0.2$
proposed by Narayan \& Medvedev (2001).  All the heated flows shown in
Figure 2 vary with time within $r \sim r_h = 25$ kpc.  
But the
negative thermal gradients produced by intermittent heating in this
region are short-lived because of the efficiency of the outward heat
flux due to thermal conduction.  Since none of the times at which the
heated flows are plotted in Figure 2 directly follows a heating
episode, the temperature gradients shown there are generally positive.

The combination of heating and thermal 
conduction also has a dramatic influence on the cooling 
rate ${\dot M}$ shown in the bottom row of Figure 2 
and in Table 1.
Of the four combinations of $\varepsilon$ and $f$ shown 
in the last two columns of Figure 2, only 
$\varepsilon = 10^{-4}$ and $f = 0.1$ has a cooling 
rate ${\dot M}$ that exceeds the 
XMM upper limits for NGC 1795 after time 4 Gyrs. 
Since we don't know
how old Abell 1795 is or when it may have 
been thermally upset by a major merger,
the age of the cluster flow must be regarded as an additional
parameter in addition to $\varepsilon$ and $f$.  
The density and
temperature profiles plotted in the second column of Figure 2 
are in satisfactory agreement with Abell 1795 observations for 
$t \lta 4$ Gyrs. 
At later
times both flows continue to agree well with the observed profiles
until $t \sim 8 - 9$ Gyrs when ${\dot M}$ exceeds the XMM upper limits
and cooling features develop just beyond the region of AGN heating at
$\sim 25 - 50$ kpc.  
Within the uncertainties, the two solutions in 
column 2 successfully match the observations of Abell 1795 for $t \lta
8$ Gyrs.

Temperature profiles in the outer regions of clusters can also
be used to constrain the thermal conduction parameter $f$,
but the observations are controversial.
Temperature profiles observed in 
cluster samples by Markevitch et al. (1998) with
{\it ASCA} and De Grandi \& Molendi (2002) with {\it BeppoSAX}
characteristically 
decline by a factor of $\sim 2$ between 0.1 and 0.5 of the
virial radius $r_{180}$.
Loeb (2002) points out that the outward
conductive flux in this region
would erase the negative temperature gradients
unless $f \lta f_{max} = 0.15 (t_{cl}/10~{\rm Gyr})^{-1}
({\bar T}/10~{\rm keV})^{-3/2}$ where 
$t_{cl}$ is the cluster age and the mean temperature
${\bar T}$ is weighted toward larger radii.
When applied to Abell 1795
(assuming $T_{max}/2 < {\bar T} < T_{max} = 6.85$ keV)
we find
$f_{max} \approx (0.26 - 0.75)(t_{cl}/10~{\rm Gyr})^{-1}$, which
is comparable to the maximum $f = 0.5$ that we consider in our
models.
However, the thermal profiles $T(r)$
determined by various detectors
and X-ray data reduction procedures are inconsistent.
For example, the XMM-PN gas temperature profile of Abell 1795
from Tamura et al. (2001) that we plot in Figures 1 and 2
continues to rise slowly at $r \sim 0.4 r_{180}$ whereas the original
combined XMM data is
nearly isothermal from 0.1 to 0.4 $r_{180}$ (Arnaud et al. 2001).
Both XMM data sets differ greatly from the mean cluster profile
of Markevitch et al. for which $dT/dr < 0$ throughout this region.
Several additional
clusters observed more recently with XMM also have nearly isothermal
thermal profiles (Arnaud et al. 2002; Pratt \& Arnaud 2002).
Since the original cluster temperature is likely to decrease
with radius (e.g. Loken et al. 2002), these nearly 
isothermal clusters clusters 
suggest that $f$ may exceed the maximum set by Loeb.

In many respects a combination of 
AGN heating with efficiency 
$\varepsilon \gta 10^{-4}$ and conductive suppression 
$f \approx 0.30 \pm 0.05$ gives excellent results for 
the evolution of rich clusters like Abell 1795.
However, Abell 1795 is far away and details of the flow 
near its central cD galaxy are not available.
In central E or cD galaxies in nearby clusters --  
M87 in Virgo, NGC 4874 in Coma and the cD in Abell 1060 -- 
the gas temperature 
drops to $\sim 1$ keV inside the optical galaxy 
(M87 with XMM: B\"ohringer et al. 2001; 
Molendi \& Gastaldello 2001;
NGC 4874 with Chandra: Vikhlinin et al. 2001;
Abell 1060 with Chandra: Yamasaki, Ohashi \& Furusho 2002). 
It is clear from the second column of Figure 2 that 
the temperature at a few kpc exceeds 1 keV, 
particularly if $\varepsilon > 10^{-4}$. 
All our 
calculations for Abell 1795 included a massive central 
E galaxy with stellar mass loss etc. so the temperature 
comparison can be made rather accurately. 
In our recent paper on M87 and NGC 4874 
(Brighenti \& Mathews 2002a) we showed that 
the sharp central temperature drop in the central E galaxy 
is possible only if $f \lta 0.1$. 
Perhaps our results in Figure 2 would improve if $f$ 
decreased at small galactic radius.

\section{Galaxy Group Flows with Conduction and Heating}

We now apply the same combination of thermal conduction 
and heating to flows on galactic scales.
For a representative X-ray  
galaxy we choose NGC 4472, a well-observed massive E1 galaxy
and the brightest galaxy in the Virgo cluster,
assumed to be at a distance of d = 17 Mpc.
NGC 4472 is also a typical 
group-centered E galaxy since it is surrounded by an extended 
cloud of hotter gas at $kT \approx 1.3$ keV, similar to 
the virial temperature of the 
galaxy group from which NGC 4472 formed.
Although the hot gas distribution in NGC 4472 is asymmetric 
at large radii, 
perhaps due to an interaction with the Virgo cluster, 
its azimuthally averaged density, shown in the top row 
of Figure 3, is similar to many other group-dominant E galaxies 
with extended X-ray emission.

Our evolutionary 
calculation for the galaxy/group differs somewhat from 
that of for the cluster.
We assume that the 
stellar configuration in NGC 4472 was assembled at some early time 
$t_{in} = 1$ Gyr in an NFW dark halo 
of total mass $M_h = 4 \times 10^{13}$ $M_{\odot}$.
The stellar density distribution $\rho_*(r)$ 
has a constant de Vaucouleurs profile 
(total mass: $M_{*t} = 7.26 \times 10^{11}$ $M_{\odot}$;
effective radius: $R_e = 1.733' = 8.57$ kpc)
with a core
$\rho_{*,core}(r) = \rho_{*,deV}(r_b)(r/r_b)^{-0.90}$
within the break radius $r_b = 2.41'' = 200$ pc
(Gebhardt 1996; Faber et al. 1997).
As the (essentially) 
single burst stellar population evolves, it expels 
mass at a rate $\alpha_* \rho_*$ gm cm$^{-3}$ s$^{-1}$ where 
$\alpha_* = 4.7 \times 10^{-20} (t/t_n)^{-1.3}$ s$^{-1}$ 
and $t_n = 13$ Gyrs is the current cosmic time. 
We assume that gas ejected from stars rapidly 
merges with the hot gas (in $\lta 10^5$ yrs). 
A modest amount of 
additional heating is provided by Type Ia supernovae 
at a rate SNu$(t) = $SNu$(t_n) (t/t_n)^{-s}$ 
where SNu$(t_n) = 0.06$ in SNu-units (supernovae per 
$10^{10}L_B$ per 100 years) gives good agreement with the 
currently observed iron abundance in the hot gas of NGC 4472. 
The circumgalactic hot gas component, required to 
reproduce the X-ray image far beyond $R_e$, is also added 
at time $t_{in}$ in hydrostatic equilibrium with the NFW halo;
we find that a variety of different scenarios for 
creating the circumgalactic gas gives similar results at time 
$t_n$.

\subsection{Unheated Galactic Flow}

As before we begin with a 2D cooling flow model for NGC 4472 
without AGN or conductive heating, but with radiative cooling. 
The evolved state of this model at time 
$t_n = 13$ Gyrs is shown in the first column of Figure 3.
The temperature distribution is in good agreement with the 
observed profile, the gas cools by radiation losses 
as it flows inward. 
However, the computed density clearly exceeds the observed 
density within about 10 kpc and fails to develop the small 
central core characteristic of E galaxies observed with Chandra
(Loewenstein et al. 2001). 
This density discrepancy is a generic difficulty with 
all simple cooling flows of this type.
The current cooling rate at the center of
NGC 4472 is ${\dot M} \approx 2.4$ $M_{\odot}$ yr$^{-1}$, 
but it was larger in the past.
Since time $t_{in} = 1$ Gyr,
a total mass $M_{cold} \approx 5.9 \times 10^{10}$
$M_{\odot}$ has cooled, but $M_{cold}$ can be
reduced somewhat if 4472 is assumed to form at a later time.
Aside from the unrealistic rise in density, the 
problems with this model are that $M_{cold}$ is too large 
(for any reasonable $t_{in}$) 
and ${\dot M}$ is probably about 5 or 10 times too large, 
based on XMM RGS observations of NGC 4636, 
an E galaxy similar to NGC 4472 (Xu et al. 2001). 

\subsection{Galactic Flows with Heating}

The second and third columns in Figure 3 show the effect  
of feedback heating for three efficiencies $\varepsilon$ 
using a Gaussian heating profile with $r_h = 2$ kpc. 
These are 2D flows, azimuthally averaged in the Figure. 
Centrally heated flows with  $\varepsilon = 10^{-5}$ 
and $10^{-4}$ (in column 2) 
differ little from the unheated flow except for 
fluctuations that accompany quasi-periodic heating episodes  
captured in the Figure at time $t_n$ during a 
short-lived excursion. 
The global properties of these two models in Table 1 are 
comparable.
The total 
cooling rate ${\dot M}$ is not appreciably reduced until
$\varepsilon$ is increased to $10^{-3}$, 
but at this level of cooling the computed density and 
(especially the) temperature 
profiles strongly disagree with observations.
The flows shown in the third column of Figure 3 are heated 
off-center with the Gaussian heating 
shifted to half the effective radius, $4.29$ kpc.
The azimuthally averaged profiles of 
these 2D flows shown at time $t_n$ in Figure 3 
have almost nothing 
in common with the observations and must be rejected.
Finally, we note that the gas cools in a larger volume
($r \lta 30$ kpc) in the heated flows (Table 1). 
This distant cooling 
arises when convective elements collide at times 
prior to a heating episode when 
the mean gas density is high; see Brighenti \& Mathews (2002b) 
for a more detailed discussion of this.

\subsection{Galactic Flows Including Thermal Conduction}

The influences of thermal conductivity and 
AGN heating on 1D flows are shown 
in the fourth and fifth columns of Figure 3.

In the fourth column of Figure 3 we show the 
influence of adding thermal conduction but with 
no additional AGN heating.
Each flow is labeled in the Figure with values of 
($\varepsilon$, $f$) also listed in Table 1. 
The 1D ($0$, $f$) solutions in column four 
fit the observed inner density profile 
at time $t_n$ but this is a momentary fluctuation. 
These solutions undergo quasi-cyclic  
transitions in $r \lta 1$ kpc 
between high density states (as in column 1 of Fig. 3) 
and the low density states shown in column 4, Figure 3.
The low density phase of the cyclic variation occurs 
just after an episode of enhanced cooling very near the 
center of the flow. 
Then the density slowly increases
toward a high state that exceeds observed densities 
in $r \lta 1$ kpc. 
These cycles do not appear in similar flows without 
thermal conduction.
The sharp rise in the gas temperature within $\sim 5$ kpc 
(column 4, Figure 3) 
is due to gravitational compression by the large 
centrally concentrated mass of 
cooled gas, which was not included in the
previously discussed 2D calculations. 
This feature is of debatable relevance and
would be much less pronounced if the galaxy formation
time $t_{in}$ were later (see below). 
Although the cooling rate ${\dot M}$ shown
in column 4 of Figure 3 is clearly reduced by conduction,
all of the cooling occurs at the
center of the flow where the cold gas mass at $t_n$ is
$M_{cold} \approx 1.5 - 2 \times 10^{10}$ $M_{\odot}$
for all $f$ considered.

These centrally cooled masses 
are unacceptably large, but most of this cooling  
occurred at very early times ($\alpha_* \propto t^{-1.3}$)
when the galactic evolution is uncertain.
For a cosmology with $H_0 = 70$ km s$^{-1}$ Mpc$^{-1}$, 
$\Omega = 0.3$ and $\Lambda = 0.7$ our initial time 
$t_{in} = 1$ Gyr corresponds to a redshift $z \approx 5$, 
somewhat larger than may be plausible.
In the fourth column of Figure 3 
we plot a third solution with $\varepsilon = 0$ and 
$f = 1$ but with the calculation starting at 
time $t_{in} = 6$ Gyrs corresponding to $z \approx 1.2$.
In this case less gas is expelled from the stars 
and a much smaller mass of gas cools at the center 
(Table 1). 
In this flow the central 
density profile exceeds observed values, similar 
to the unheated flow without conduction in the first column.

In the final column of Figure 3 we show two flows 
with both AGN heating and thermal conduction.
The flow with $\varepsilon = 10^{-3}$ and $f = 0.25$
is an excellent fit to the density and temperature profiles, 
disregarding the central spike in the temperature 
created as gas is compressed by 
the point-like gravitational attraction of centrally cooled gas.
But the global cooling rate for this solution,
${\dot M} \approx 1.6$ $M_{\odot}$ yr$^{-1}$, 
is too large to be consistent with typical XMM spectral limits.
Another solution of this type with 
$\varepsilon = 5 \times 10^{-3}$ and $f = 1$
is seen to have a good density profile and a very low ${\dot M}$, 
but the temperature is clearly too high for $r \lta 10$ kpc. 
This temperature rise is an important defect in the flow 
and one that cannot be removed by fine-tuning $\varepsilon$ 
and $f$. 
Because of the lower temperature and conductivity 
$\kappa \propto T^{5/2}$ in galaxy/group flows, 
the conduction needs to be much closer to the Spitzer value 
in order to achieve results similar to the cluster flows.

Another quite different problem with these marginally  
acceptable galaxy/group flows is the iron abundance 
in the hot gas. 
Typically the iron abundance is $\sim$solar near the 
cores of group-centered E galaxies
(e.g. Buote et al. 2002) where the gas is enriched 
by Type Ia supernovae, each 
producing $\sim 0.7$ $M_{\odot}$ of iron.
At least some of this iron must go into the hot gas phase 
since the iron abundance in the gas increases toward the center of 
the stellar system. 
However in otherwise acceptable 
flows such as those in the last two columns of Figure 3, 
where the gas density approximately agrees with observations,
our computed iron abundance is much larger than observed.  
The iron abundance at $t_n$ is shown in Figure 4 for the flow with 
$\varepsilon = 5 \times 10^{-3}$ and $f = 1$. 
For example, the iron abundance in the computed flow greatly exceeds 
that observed by Buote (2000).
One way to avoid this problem, and also retain some of the 
good features of the heating plus conduction solutions, 
may be to carefully regulate 
(i.e., fine-tune) the fraction of Type Ia iron 
that ultimately goes into the hot gas. 
Alternatively, the gas may be cooling in some fashion that 
cannot be observed with XMM.

\section{Final Remarks and Conclusions}

Cluster flows computed
with both AGN heating and thermal conduction
reduced by a factor $0.35 \pm 0.10$ from the Spitzer value
can provide acceptable fits to cluster flow observations.
Within broad limits, 
successful flows are insensitive to the (Gaussian) width 
$r_h$ of the feedback-heated region. 
The observed density and temperature profiles can be maintained
with lower cooling rates as required by XMM spectra.
Nevertheless, in these solutions the gas temperature may be too
high within $\sim 10$ kpc
of the center, i.e. inside the central cD galaxy.
If so, it may be necessary to consider flows with
a radially variable conductive suppression factor $f$.

Flows on galactic scales also benefit from the combined
effects of AGN heating and conduction, but the agreement
is less satisfactory than for cluster flows.
The parameters that provide marginally acceptable, but not ideal,
flow solutions on galaxy/group scales -- ($\varepsilon$,$f$) =
(0,0.5),(0,1) and ($5 \times 10^{-3}$,1) -- require that $f$ be much
closer to the Spitzer value than for cluster flows.  Such large values
$f \approx 0.5 - 1$ -- at the higher temperatures of clusters -- would
be inconsistent with the ``cold fronts'' observed in clusters
(e.g. Markevitch et al. 2000) and may promote cluster 
isothermality as described by Loeb (2002).  Values of $f \gta 0.5$
also exceed the predictions of Narayan \& Medvedev (2001).  Perhaps
the field geometry is more favorable to thermal conduction in
undisturbed parts of the flows, which we emphasize here, than in the
cold fronts where fields transverse to the thermal gradient are
expected (Ettori \& Fabian 2000).
Nevertheless, higher values of the conductivity coefficient
$f$ are required in galaxies/groups to achieve the same 
benefits as in hotter cluster gas and it can be 
doubted if such large conductivities are physically acceptable. 

Another difficulty with our galaxy/group heated flows 
are the high hot gas iron abundances, 
several times higher than observed.
The high iron abundance occurs in the computed flows
when the gas density is lowered (by heating) to fit the
observed profile.
The Type Ia supernova rate that we assume,
SNu$(t_n) = 0.06$ SNu, is already much lower than the
combined rate observed in E and S0 galaxies,
$(0.16 \pm 0.05)h_{70}^2$ (Cappellaro et al. 1999). 
Therefore, to fully accept the beneficial effects of 
heating plus conduction in galactic flows, 
it is also necessary to hypothesize some means of reducing 
the computed iron abundance: some (but not all of)
the iron produced in Type Ia may cool before entering
the hot gas, some distributed cooling is present to 
remove the excess iron, etc.
Flows on the galaxy/group scale are much more
constrained by X-ray observations than cluster scale flows,
although the latter have received by far the most attention
from observers.

%\end{document}

\vskip.4in
Studies of the evolution of hot gas in elliptical galaxies
at UC Santa Cruz are supported by
NASA grant NAG 5-8049 and NSF grants  
AST-9802994 and AST-0098351 for which we are very grateful.
FB is supported in part by grant MURST-Cofin 00.

%\clearpage
\vskip.1in
\figcaption[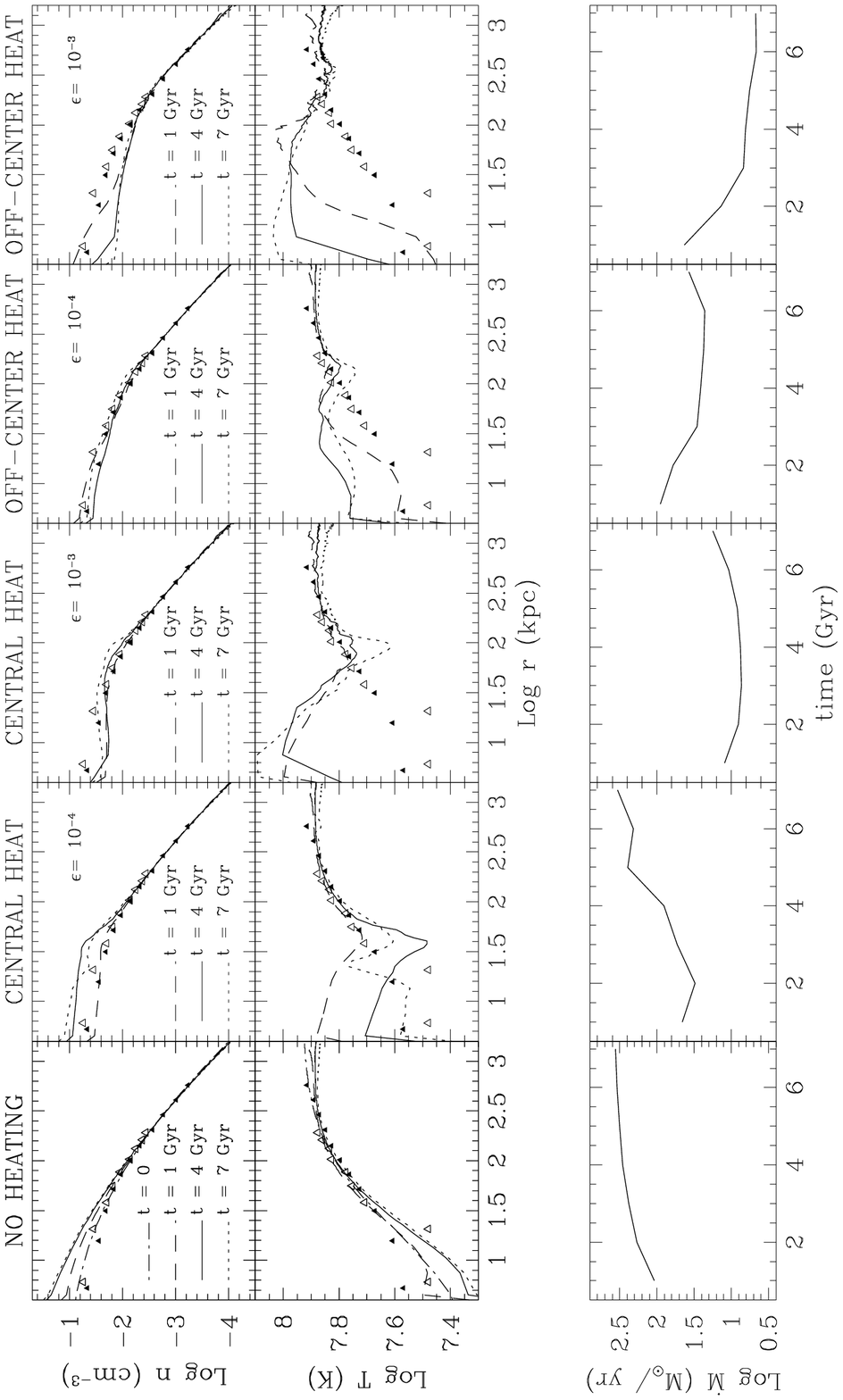]{
Evolution of cluster flows modeled after Abell 1795.
Observations of 
the hot gas electron density and temperature are shown 
as a function of radius in the first two rows:
XMM ({\it filled triangles}; Tamura et al. 2001)
and Chandra ({\it open triangles}; Ettori et al. 2002).
All models are 2D and 
begin in hydrostatic equilibrium based on 
the fits to $n(r)$ and $T(r)$ shown as {\it dot-dashed lines} in the 
left column.\\
The bottom row shows the variation of the total 
radiative cooling rate as a function of time. 
The first column shows the evolution of the hot 
gas after evolving for $t = 0$ Gyr ({\it dot-dashed line}), 
1 Gyr ({\it dashed line}), 
4 Gyrs ({\it solid line}), and
7 Gyrs ({\it dotted line}).
The second and third columns show the evolution of 
centrally heated flows 
with AGN efficiencies $\varepsilon = 10^{-4}$ 
and $10^{-3}$ respectively.
The fourth and fifth columns show the evolution of 
flows that are heated 50 kpc from the center 
with efficiencies $\varepsilon = 10^{-4}$
and $10^{-3}$ respectively.
All computed densities and temperatures are azimuthally averaged 
and shown after 1, 4 and 7 Gyrs with the same 
line types as in the first column.
\label{fig1}}

\vskip.1in
\figcaption[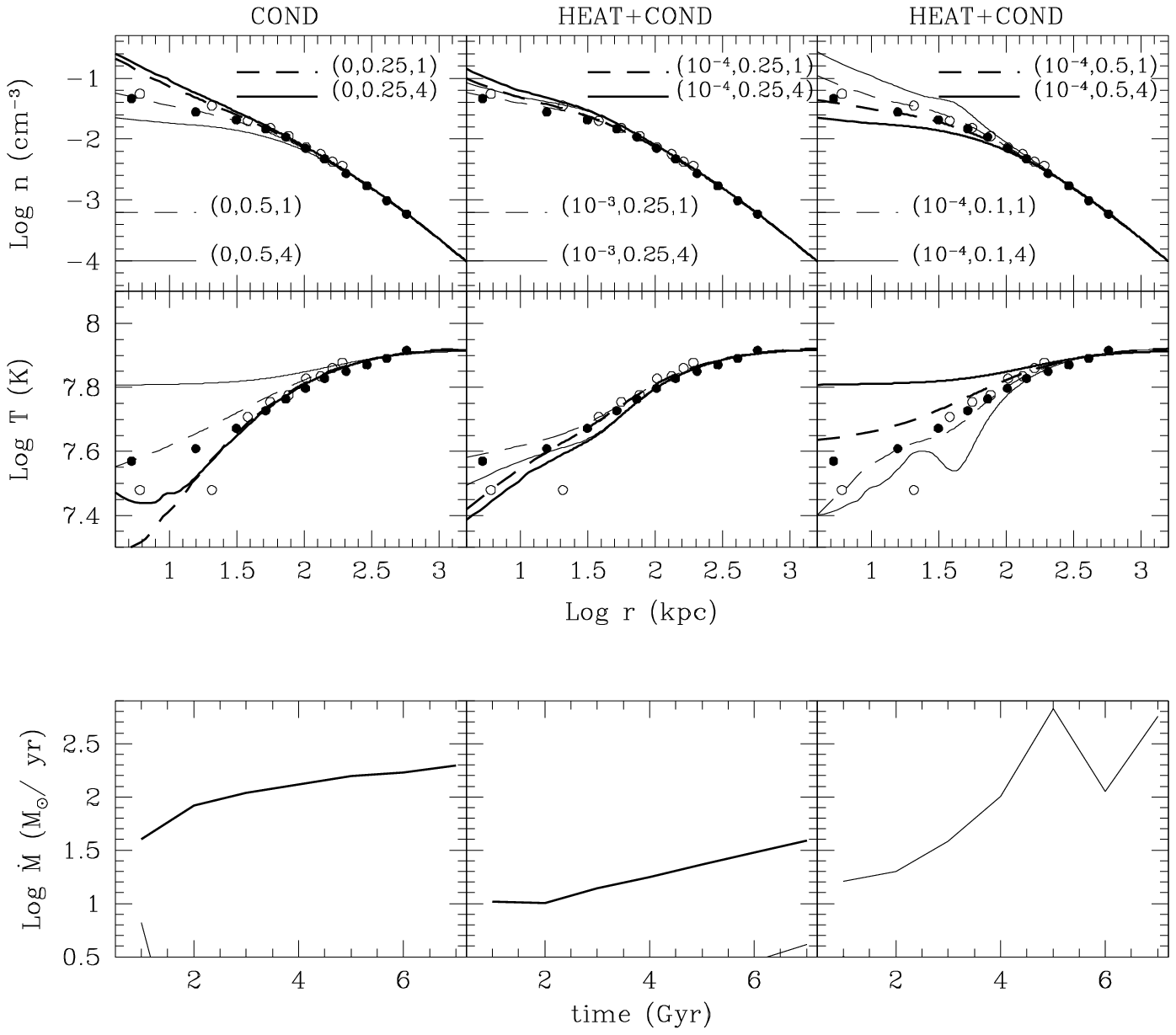]{
Evolution of cluster flows modeled after Abell 1795 
including thermal conduction.
All models are 1D and
begin in hydrostatic equilibrium.
In descending order the rows show the computed 
density $n(r)$ and temperature $T(r)$ variations and the 
third row shows the total cooling rate as a function of time.
Each curve is characterized by three numbers 
($\varepsilon$,$f$,$t$) where 
$\varepsilon$ is the feedback heating efficiency, 
$f$ is the fraction of the Spitzer conductivity used 
and $t$ is the time in Gyrs since the calculation began.
The observations are identical to those in Figure 1.
\label{fig2}}

\vskip.1in
\figcaption[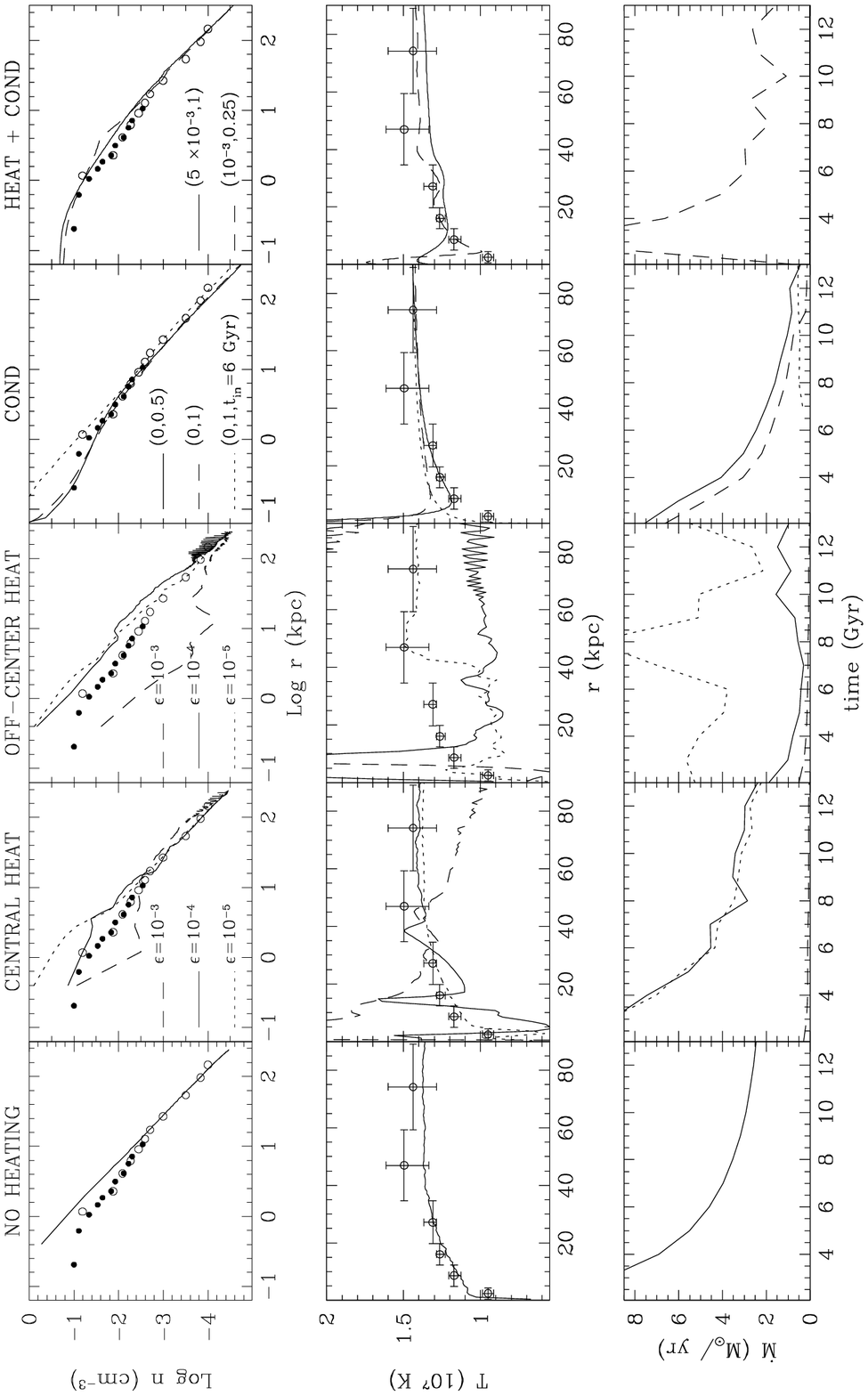]{
Evolution of galactic/group flows shown at time 
$t_n = 13$ Gyrs. 
In descending order each row shows the density profile 
$n(r)$, the temperature profile $T(r)$ and the variation of 
the total cooling rate with time during the calculation.
The temperatures are shown as functions of physical radius
and are azimuthally averaged. 
Density observations for NGC 4472
are from {\it Einstein} (Trinchieri, Fabbiano,
\& Canizares 1986) ({\it filled circles}) and
ROSAT (Irwin \& Sarazin 1996) ({\it open circles}), 
the temperature observations are from Irwin \& Sarazin (1996).\\
The first column shows a 2D cooling flow solution with no 
AGN heating or thermal conduction.
The next two columns illustrate 2D flows with feedback heating 
at the center and flows 
with off-center heating at $R_e/2$. 
Each computed model is labeled with the feedback efficiency 
$\varepsilon$ used.
In the solution shown with {\it short dashed lines} in column 4 
the calculation was begun 
at $t_{in} = 6$ Gyrs not 1 Gyr as in all other models.
The last two rows are 1D flows for thermal conduction only and 
conduction plus AGN heating respectively. 
The pair of parameters ($\varepsilon$, $f$) apply to 
each particular solution. 
\label{fig3}}

\vskip.1in
\figcaption[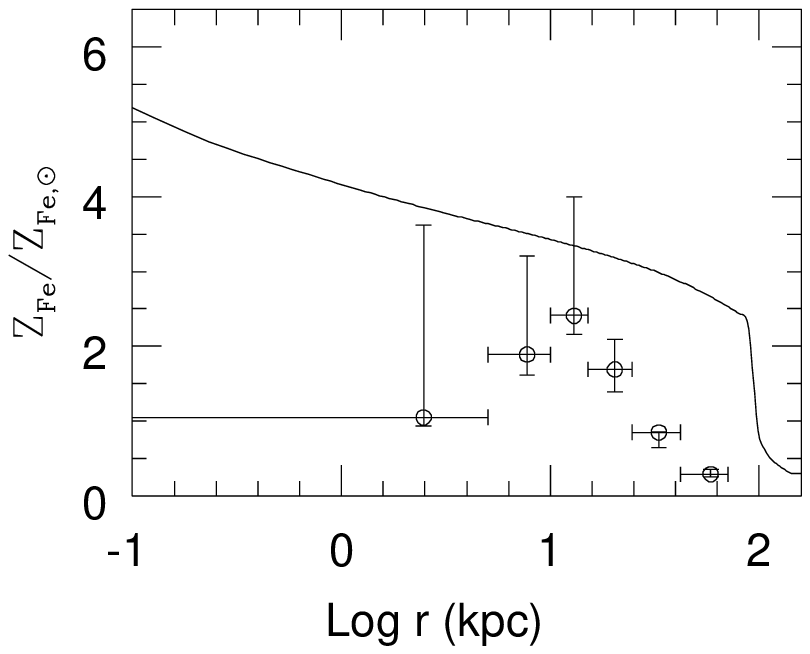]{
Iron abundance in the hot gas at $t_n = 13$ Gyrs for 
the galactic/group flow with $\varepsilon = 5 \times 10^{-3}$ 
and $f = 1$. 
The points are the iron abundance observed in NGC 4472 
by Buote (2000). 
\label{fig4}}

\newpage

\makeatletter
\def\jnl@aj{AJ}
\ifx\revtex@jnl\jnl@aj\let\tablebreak=\nl\fi
\makeatother

\begin{deluxetable}{lcccrcrcccc}
\scriptsize
%\rotate
\tablewidth{18cm}
%\tablenum{1}
%\tablewidth{0pt}
\tablecolumns{10}
\tablecaption{PARAMETERS AND SUCCESS OF CLUSTER AND GALACTIC FLOWS}
\tablehead{
\colhead{Flow} &
\colhead{1D} &
\colhead{$\varepsilon$\tablenotemark{a}} &
\colhead{heating\tablenotemark{b}} &
\colhead{$r_h$\tablenotemark{c}} &
\colhead{$f$\tablenotemark{d}} &
\colhead{${\dot M}$\tablenotemark{e}} &
\colhead{$M_{cold}(0)$\tablenotemark{f}} &
\colhead{$M_{cold}$;$\langle r \rangle$\tablenotemark{g}} &
\colhead{Problem\tablenotemark{h}} &
\colhead{Verdict\tablenotemark{i}} \\
\colhead{} &
\colhead{or 2D} &
\colhead{} &
\colhead{} &
\colhead{(kpc)} &
\colhead{} &
\colhead{($M_{\odot}$/yr)} &
\colhead{($10^{10}M_{\odot}$)} &
\colhead{($10^{10}M_{\odot}$; kpc)} &
\colhead{} &
\colhead{}
}
\startdata
Cluster  & 2D & 0 & N & ... & 0 & 289 & 83 & 83; 0 & ${\dot M}$&R \\
 & 2D & $10^{-4}$ &         C &  25 & 0 &  80 & 21 & 21; 0 & ...& A? \\
 & 2D & $10^{-3}$ &         C &  25 & 0 & 7.5 &3.5 &3.5; 0 &$\rho$,$T$&R \\
 & 2D & $10^{-4}$ &        OC &  25 & 0 &  26 & 20 & 20; 0 &$T$? &R \\
 & 2D & $10^{-3}$ &        OC &  25 & 0 &  4.7  &  7 &  7; 0 &$\rho$,$T$& R \\
 &    &           &           &     &   &     &    &       &  & \\
 & 1D & 0         & N & ... & 0.25 & 131 &  36 & 36; 0 &$\rho$,${\dot M}$ &R \\
 & 1D & 0         & N & ... & 0.35 &  19 & 8.4 &8.4; 0 &... & A\tablenotemark{j} \\
 & 1D & 0         & N & ... & 0.5  &   0 & 0.7 &0.7; 0 &$\rho$,$T$& R \\
 & 2D & $10^{-4}$ &         C &  25 & 0.25 &  18 &  5 & 5; 0 &...& G \\
 & 2D & $10^{-3}$ &         C &  25 & 0.25 &  2 & 0.6 & 0.6; 0 &...& G \\
 & 2D & $10^{-4}$ &         C &  25 & 0.1  &  100 & 17 & 17; 0 &$\rho$,${\dot M}$& R \\
 & 2D & $10^{-4}$ &         C &  25 & 0.5 &  0  &  0.3 & 0.3; 0 &$\rho$,$T$& R \\
 &    &           &           &     &   &     &    &       &   &\\
Galaxy  & 2D & 0 & N & ... & 0 & 2.4 & 5.9 & 5.9; 0 &$\rho$,${\dot M}$ &R \\
 & 2D & $10^{-5}$ &         C &   2 & 0 &  2.2 & 4.6 & 5.9; 4 &$\rho$,${\dot M}$& R \\
 & 2D & $10^{-4}$ &         C &   2 & 0 &  2.4 & 1.2 & 5.9;12 &$\rho$,${\dot M}$& R \\
 & 2D & $10^{-3}$ &         C &   2 & 0 & 0.05 & 0.1 & 0.1; 0 &$\rho$,$T$& R \\
 & 2D & $10^{-5}$ &        OC &   2 & 0 &  5.4\tablenotemark{k} & 4.8 & 6.1;30 &$\rho$,$T$,${\dot M}$& R \\
 & 2D & $10^{-4}$ &        OC &   2 & 0 &  1.0 & 1.0 & 1.1;10 &$\rho$,$T$,${\dot M}$& R \\
 & 2D & $10^{-3}$ &        OC &   2 & 0 & 0.04 & 0.2 & 0.2; 0 &$\rho$,$T$& R \\
 & 1D &    0      &         N & ... &0.5 &  0.4 & 1.9 & 3.1; 1 &${\dot M}$?,$z_{Fe}$& A? \\
 & 1D &    0      &         N & ... & 1 & 0.14 & 1.4 & 2.3;0.8&$T$?,$z_{Fe}$& R? \\
 & 1D\tablenotemark{l} &    0      &         N & ... & 1 & 0.54 & 0.2 & 0.3;0.5&$T$?& A? \\
 & 1D & $10^{-3}$ &        C &   2 & 0.25 & 1.6 & 0.017 & 4.2;5 &${\dot M}$,$z_{Fe}$& R \\
 & 1D & $5 \times 10^{-3}$ &C&   2 & 1 &  0.002 & 0.004 & 0.004;0 &$T$?,$z_{Fe}$& A? \\

%\enddata
\tablenotetext{a}{AGN heating efficiency.}
\tablenotetext{b}{Location of heating: C (center) or OC (off-center) 
or N (none).}
\tablenotetext{c}{Gaussian scale of heated region.}
\tablenotetext{d}{Fraction of Spitzer conductivity.}
\tablenotetext{e}{Total cooling rate at 4 Gyrs (cluster flows) or 
13 Gyrs (galactic flows).}
\tablenotetext{f}{Total mass of cooled gas at the origin 
at 4 Gyrs (cluster flows) or 13 Gyrs (galactic flows).}
\tablenotetext{g}{Total mass of cooled gas and approximate radius 
within which the cooling occurs 
at 4 Gyrs (cluster flows) or 13 Gyrs (galactic flows).}
\tablenotetext{h}{This column lists the main problem with the computed flow.}
\tablenotetext{i}{Final qualitative judgment on the quality of the 
flow: R (rejected); A (acceptable); G (good).}
\tablenotetext{j}{Not plotted in Figure 2.}
\tablenotetext{k}{This high ${\dot M}$ is a momentary fluctuation, the
average value is $\sim 2$.}
\tablenotetext{l}{For this flow $t_{in} = 6$ Gyrs.}
\enddata
\end{deluxetable}

\end{document}